\documentclass{PoS}
\usepackage{epsfig}

\PoS{PoS(LAT2005)319}

\title{String tensions and deconfinement transition in the SU(4)
center vortex model }

\ShortTitle{SU(4) center vortex model }

\author{\speaker{Michael Engelhardt}\thanks{Supported by the U.S. DOE
under grant number DE-FG03-95ER40965.}\\
New Mexico State University\\
E-mail: \email{engel@nmsu.edu}}

\abstract{A center vortex model for the infrared sector of $SU(4)$ Yang-Mills
theory is constructed such as to reproduce both the ratio between the
zero-temperature quark and diquark string tensions known from lattice
Yang-Mills theory, as well as the properties of the deconfinement transition.
On this basis, the temperature dependence of the spatial quark and diquark
string tensions is predicted. Though still phenomenologically
viable, details of the construction of the $SU(4)$ center vortex model
corroborate previous arguments that modeling infrared Yang-Mills
dynamics purely in terms of vortex world-surface characteristics may
become less appropriate as the number of colors is increased.}

\FullConference{XXIIIrd International Symposium on Lattice Field Theory\\
		 25-30 July 2005\\
		 Trinity College, Dublin, Ireland}

\begin{document}

\section{Introduction and Motivation}
Random vortex world-surface models are designed to describe the infrared
sectors of $SU(N)$ Yang-Mills theories. They are based on the notion that the
relevant infrared degrees of freedom are center vortices, i.e., closed tubes
of quantized chromomagnetic flux
\cite{hooft,mack,olesen,jeff,percol,philippe}. In practice, ensembles of
center vortex world-surface configurations are generated using Monte Carlo
methods, where the world-surfaces are composed of elementary squares on a
hypercubic lattice.

In the case of $SU(2)$ color, such a vortex model has proven to furnish
a comprehensive description of the central mechanisms governing the strong
interaction in the infrared \cite{su2conf,su2top,su2csb}. Both the confined
and deconfined phases of the corresponding Yang-Mills theory are generated,
separated by a second order finite-temperature deconfinement transition. In
addition to the confining properties, also the topological properties encoded
in the topological susceptibility are reproduced quantitatively; furthermore,
the (quenched) chiral condensate in the vortex ensemble is quantitatively
compatible with the behavior found in $SU(2)$ Yang-Mills theory.

In the $SU(3)$ case, the random vortex world-surface model shows similar
promise \cite{su3conf,su3bary,su3freee}. While the topological properties and
the coupling to quark degrees of freedom remain to be investigated, the
$SU(3)$ vortex model again generates both the confined and deconfined phases,
separated by a weak first order deconfinement transition. The static baryonic
potential satisfies a Y law and the vortex free energy displays the
correct behavior as a dual order parameter.

In both the $SU(2)$ and $SU(3)$ models, only one dimensionless parameter
characterizing the vortex world-surface dynamics needs to be adjusted
to arrive at the quantitative agreement with lattice Yang-Mills theory
highlighted above. This parameter controls vortex stiffness,
i.e., it is the coefficient of a curvature action term. On the other hand,
this simple picture, with dynamics controlled purely by one vortex
world-surface characteristic, is not expected to persist for a larger
number of colors, as was previously argued in \cite{jeffstef}.
This can be understood as follows: Generic vortex world-surfaces are not
orientable, which implies that they cannot be described by a field strength
which points into a single direction in color space. Rather, the color
direction of the field strength must have some amount of variation as one
varies position on the vortex, in any gauge. In particular, for many
purposes it is convenient to construct the vortex field strength in an
Abelian fashion, such that it is restricted to a definite (Cartan) color
component for each possible type of vortex flux.
In this case, the field strength in general is forced to jump at
certain lines defined on the vortex world-surface in a way which corresponds
to having a source or sink of magnetic flux at those lines. In other words,
these lines constitute the world-lines of Abelian monopoles, which are
thus an intrinsic feature of generic, non-orientable, vortex world-surface
configurations in Abelian gauges \cite{cornwall,contvort,su2top}. Moreover,
in the case of three or more colors, these monopoles appear not only as
locations at which the field strength on a vortex jumps, but the vortex
flux may also branch there; in general, for $N$ colors, up to
$N$ center vortex fluxes can emanate from any given monopole.

\begin{figure}[h]
\centerline{\epsfig{file=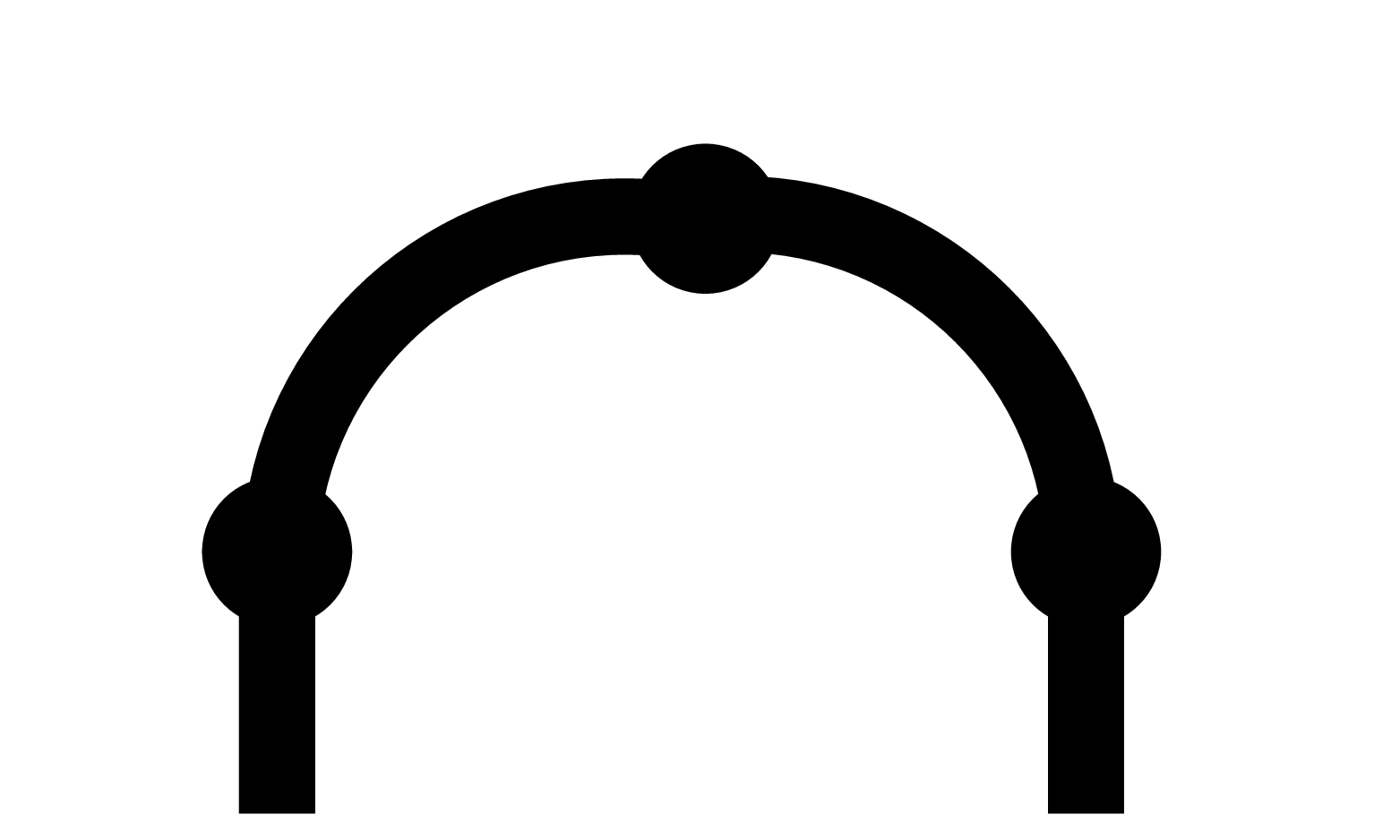,width=4cm} \hspace{2cm}
\epsfig{file=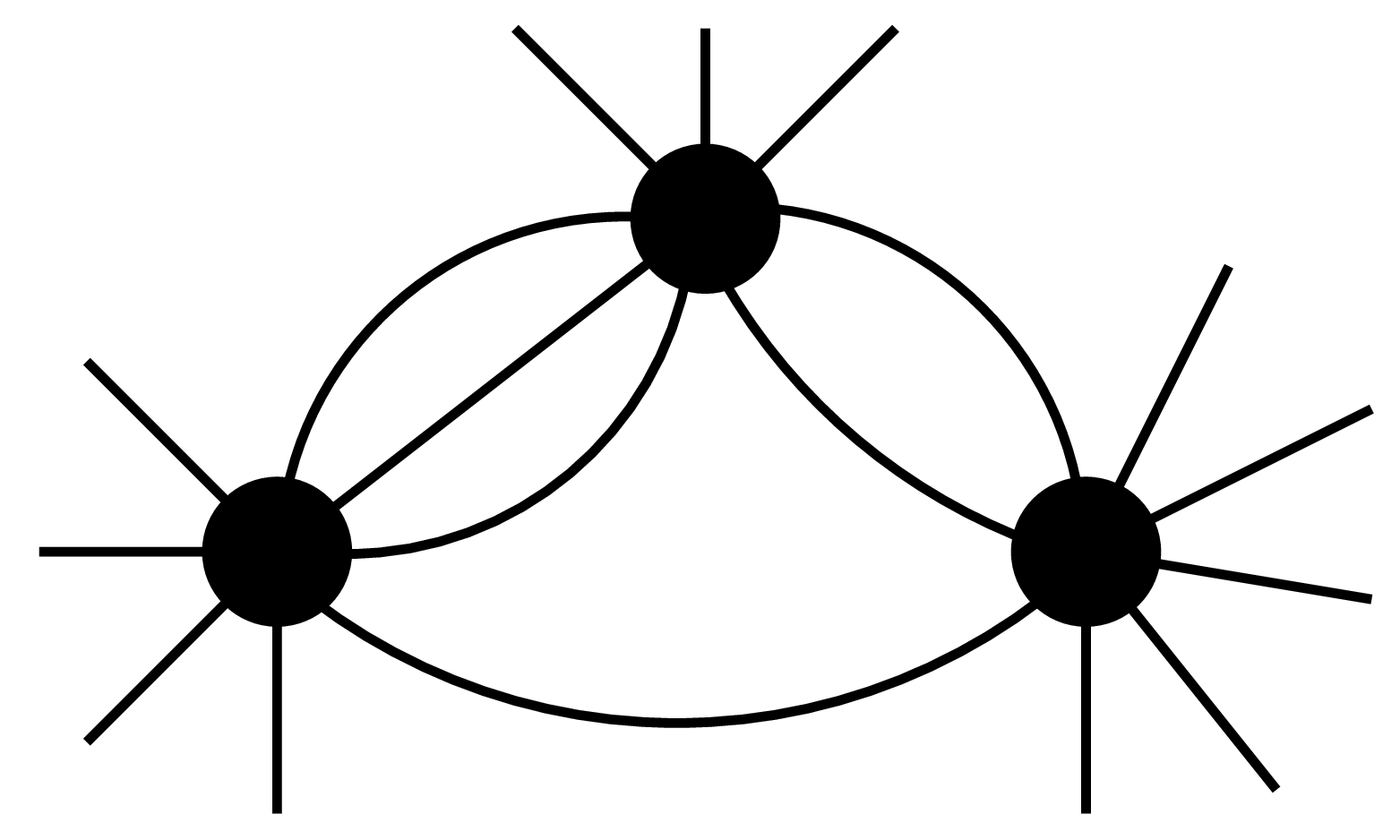,width=4cm}}
\caption{Vortex configurations contain monopoles. For $N$ colors, up to
$N$ vortex fluxes can emanate from any given monopole. Examples for $N=2$
(left) and $N=8$ (right) are depicted. As $N$ rises, center flux can be
quantized in smaller units, while monopoles remain sources or sinks of
the same total amount of flux.}
\label{monopfig}
\end{figure}

As the number of colors $N$ rises, center flux can be quantized in ever
smaller units, while the monopoles always constitute sources or sinks
of the same total amount of flux, cf.~Fig.~\ref{monopfig}. It seems
plausible that this may lead to a shift in the relative importance of pure
vortex world-surface characteristics and monopole characteristics for the
dynamics of the vortex ensemble. While at low $N$, the monopole dynamics
appear to be completely determined by the dynamics of the vortices on which
the monopoles reside, at higher $N$, the monopoles may attain their own
dynamic significance and may influence the overall dynamics of the vortex
world-surfaces. Thus, a pure vortex world-surface curvature action, which
appears to be entirely sufficient to model $SU(2)$ and $SU(3)$ Yang-Mills
theory in the infrared regime, may well cease to be appropriate at higher
$N$. Definite indications of this occur in the $SU(4)$ case investigated
in the present work, which thus aims to probe the limits of applicability
of a pure random world-surface dynamics for center vortices.

\section{SU(4) vortex model}
The $SU(4)$ gauge group contains the four center elements $\{ 1,i,-1,-i\} $,
which determine the quantization of vortex flux; vortices are
characterized by contributing a center phase to any Wilson loop to which
they are linked. The trivial unit center element corresponds to no vortex
flux being present, and vortices generating the phases $i$ and $-i$ are
related by a space-time inversion. Therefore, there are actually two
different types of vortices; the ones generating a $-1$ phase and the ones
generating the phases $\pm i$ depending on their orientation in space-time.

Correspondingly, $SU(4)$ Yang-Mills theory contains two
different string tensions, the quark string tension $\sigma_{1} $ and
the diquark string tension $\sigma_{2} $. Contrary to the $SU(2)$ and
$SU(3)$ cases, where one string tension measurement is needed as input
to fix the stiffness of one vortex type, for $SU(4)$, measurements of
both string tensions, $\sigma_{1} $ and $\sigma_{2} $, are needed as input
to fix the stiffness of the two vortex types in $SU(4)$.
In practice, the two pieces of data from lattice Yang-Mills ``experiment''
used to fix the curvature action are ($T_C $ denoting the
deconfinement temperature) \cite{luctep03,luctep04}
\begin{equation}
\sigma_{2} (T=0) / \sigma_{1} (T=0) = 1.36
\ \ \ \ \ \ \ \ \ \ \ \ \ \ \ \ \ \ 
T_C / \sqrt{\sigma_{1} (T=0)} = 0.62
\label{input1}
\end{equation}

A $SU(4)$ vortex action using two stiffness parameters for the two types
of vortices would constitute the direct generalization of the actions used
in the $SU(2)$ and $SU(3)$ cases. However, initial investigation of such
a model, adjusting two stiffness parameters to the inputs (\ref{input1}),
led to the conclusion that it is not phenomenologically viable.
Specifically, the deconfinement transition, if at all first order, turned
out to have an undetectably low latent heat, certainly considerably lower
than found in the $SU(3)$ vortex model. This is in qualitative
disagreement with $SU(4)$ lattice Yang-Mills theory, where the
transition is roughly twice as strongly first order \cite{luctep03}
as for $SU(3)$. Thus, as detailed below, an additional action term
favoring vortex branching was introduced, and its
strength was fixed using a third input from lattice Yang-Mills ``experiment'',
characterizing the deconfinement transition \cite{luctep03}:
\begin{equation}
\left. \Delta S \frac{T_C^4 }{\sigma_{1}^{4} (T=0)} \right|_{SU(4)} = 
2\cdot \left. \Delta S \frac{T_C^4 }{\sigma^{4} (T=0)} \right|_{SU(3)}
\label{input2}
\end{equation}
where $\Delta S$ denotes the latent heat.
In practice, the center vortex world-surfaces are modeled by composing them
of elementary squares on a hypercubic lattice. The lattice spacing in
this approach is a fixed physical quantity implementing the notion that
vortices possess a finite transverse thickness and must be a minimal 
distance apart to be distinguished from one another.
Vortex world-surfaces on lattices are generated using Monte Carlo methods.
The action can be represented as
\vspace{0.5cm}
\begin{equation}
\hspace{-2cm} S\ \ = \ \ c_i \, \, c_j \ \times
\hspace{1.5cm} - \ \ b \ \times
\label{voract}
\end{equation}
\vspace{-2cm}

\begin{figure}[h]
\centerline{\hspace{1.6cm} \epsfig{file=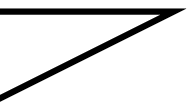,width=1.7cm}
\hspace{0.6cm} \epsfig{file=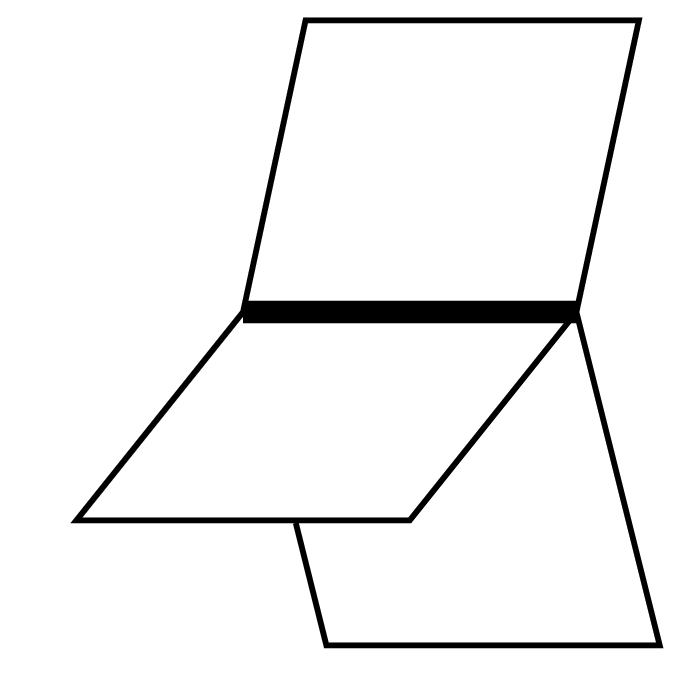,width=1.53cm} }
\end{figure}

\vspace{-2.3cm}

\hspace{6.3cm} $j$

\vspace{0.95cm}

\hspace{5.9cm} $i$

The first term is the curvature term; it penalizes configurations in which
two vortex squares share a link without lying in the same plane. This
directly generalizes the action used in the $SU(2)$ and
$SU(3)$ cases. There, only one type of vortex exists, and there is only
one curvature coefficient weighting vortex world-surfaces ``going around
a corner''. In the present case, the strength of the curvature action
varies according to the different possible types of vortex squares meeting
at a link. Note that it would be quite admissible to use a more general
curvature coefficient $c_{ij} $ with $c_{12} \neq c_1 \, c_2 $. This
option was explored, but did not seem to enhance the phenomenological
flexibility of the model. The same is true for an area action term which
simply penalizes the total vortex surface area, the effects of which
can in practice be absorbed into the curvature term \cite{su2conf}.

The second term in the action, on the other hand, encourages vortex
branching: The action is decremented by the branching
coefficient $b$ for each link which has 3 or 5 vortex squares attached
to it. This happens precisely when a vortex branches. As indicated above,
this term is introduced in order to enhance the first order
character of the deconfinement transition. The underlying rationale
is this: The $SU(2)$ and
$SU(3)$ models are governed by the same type of action; they differ
in the topology of the configurations. Only the $SU(3)$ model
contains vortex branchings and only it exhibits a first order deconfinement
transition. Thus, it appears plausible that enhancing branchings in the
$SU(4)$ model may restore a detectable first order deconfinement transition
of the correct strength, cf.~(\ref{input2}). This is indeed confirmed by
the measurements described below. Before proceeding to discuss
these, it should be noted that vortex branchings are associated with
the presence of a monopole if one specifies the vortex field strength
such that it is restricted to a definite (Cartan) color component for
each type of vortex flux. Therefore, the branching coefficient $b$ can
also be interpreted as favoring the presence of monopoles. In this sense,
the action (\ref{voract}) accords monopoles their own dynamical
significance and ceases to rely purely on
vortex world-surface characteristics.

In practice, the physical point of the model was found as follows. Since
only discrete temperatures can be realized using a fixed lattice spacing,
at the physical point in parameter space the deconfinement transition is
in general not directly accessible. Instead, the physical point is fixed by
interpolation of parameter sets at which an integer multiple of the
lattice spacing indeed corresponds to the inverse deconfinement temperature.
In practice, measurements were carried out on lattices with $N_t =1,2,3$
spacings in the temporal direction, and the coupling constants $c_1 $,
$c_2 $ and $b$ were fixed such that the deconfinement temperature is
realized, with the first condition in (\ref{input1}) and the
condition (\ref{input2}) satisfied as well. The result is displayed in
Table \ref{physpt}, along with the ratio $T_C /\sqrt{\sigma_{1} } $
obtained at each parameter set.
The physical point is found from these data by interpolating quadratic
functions of $T_C /\sqrt{\sigma_{1} } $ for the three coupling constants.
Setting $T_C /\sqrt{\sigma_{1} } =0.62$,
which is the final condition to be satisfied at the physical
point, cf.~(\ref{input1}), defines the physical parameters
\pagebreak
\begin{equation}
c_1 = 0.45 \ \ \ \ \ \ \ c_2 = 0.80 \ \ \ \ \ \ \ b=0.71
\label{physparms}
\end{equation}
As is evident from Table \ref{physpt}, the physical point is very near
the parameters obtained for $N_t =2$; the uncertainties inherent in the
interpolation thus remain small. As a cross-check, it was verified
that the first condition in (\ref{input1}) remains satisfied for the
parameters (\ref{physparms}).

\begin{table}[h]
\begin{center}
\begin{tabular}{|c||c|c|c||c|}
\hline
& $c_1 $ & $c_2 $ & $b$ & $T_C /\sqrt{\sigma_{1} } $ \\
\hline \hline
$N_t =1$ & 0.2785 & 0.4005 & 0.1403 & 0.90 \\
\hline
$N_t =2$ & 0.4558 & 0.7983 & 0.6950 & 0.61 \\
\hline
$N_t =3$ & 0.5925 & 0.7059 & 0.3800 & 0.50\\
\hline
\end{tabular}
\end{center}
\caption{Sets of coupling constants realizing the deconfinement temperature
as well as satisfying two of the three conditions defining the physical
point. The last condition, $T_C /\sqrt{\sigma_{1} } =0.62$, is satisfied
by interpolating the parameters using the data in the final column.}
\label{physpt}
\end{table}

Having found the physical point, the behavior of the string tensions
as a function of temperature can be predicted. In particular, the
spatial string tensions $\sigma_{1}^{S} $ and $\sigma_{2}^{S} $ in the
deconfined phase are of interest. Using a lattice with $N_t =1$ realizes
the temperature $T=1.94 T_C $, and one obtains
$\sigma_{1}^{S} (T=1.94 T_C ) = 1.34 \ \sigma_{1} (T=0)$ and
$\sigma_{2}^{S} (T=1.94 T_C ) = 1.44 \ \sigma_{2} (T=0)$.
The characteristic rise of the spatial string tensions in the deconfined
phase is observed; more specifically, this rise is such that the ratio
$\sigma_{2}^{S} /\sigma_{1}^{S} (T=1.94 T_C ) = 1.46 $ is enhanced
compared to the zero-temperature value 1.36. For comparison, if one considers
the simplified model with no branching term, $b=0$, and in turn disregards
condition (\ref{input2}), this ratio takes the value 1.25, smaller than
the zero-temperature value.

\section{Conclusions}
A description of the infrared sector of $SU(4)$ Yang-Mills theory in terms
of a random vortex world-surface ensemble is
viable; however, the effortless predictivity of the $SU(2)$ and $SU(3)$
models is lost. In those cases, it proved sufficient to base the
dynamics on a single vortex world-surface characteristic, namely the
curvature. By contrast, in the $SU(4)$ case, it is not only necessary to
introduce two different curvature coefficients due to the existence of two
different vortex types; in addition, a description based purely on
world-surface properties had to be abandoned in favor of an action
which also endows monopoles (which are intrinsically present in vortex
configurations)
with their own dynamical significance. This corroborates related arguments
put forward in \cite{jeffstef}.

Three dimensionless parameters had to be tuned in order to
correctly reproduce the properties (\ref{input1}) and (\ref{input2}).
The model thus seems rather less attractive in terms
of predictive power than the $SU(2)$ and $SU(3)$ models. Nevertheless,
the behavior of the spatial string tensions in the deconfined phase was
predicted. During the conference, the author became aware of newer
lattice Yang-Mills data to which these quantities can be compared
\cite{luctepnew}. According to these data, the ratio of the (spatial)
diquark and quark string tensions remains very near its zero-temperature
value $\sigma_{2} /\sigma_{1} =1.36$ at high temperatures.
Comparing with the results quoted in the previous section,
introducing the branching term into the action (\ref{voract}) thus
shifts the value of $\sigma_{2} /\sigma_{1} $ at high temperatures in
the correct direction compared with the $b=0$ model; however, the
correction overshoots the lattice Yang-Mills data considerably. Presumably,
also this could be adjusted, e.g., by introducing an additional free
parameter $c_{12} \neq c_1 \cdot c_2 $ in place of the combination
$c_1 \cdot c_2 $ into the action (\ref{voract}). Without going into further
detail concerning such generalized models, this discussion does clearly
underscore once more the observation made at the beginning of this section,
that the effortless predictivity of the $SU(2)$ and $SU(3)$ cases is lost
in the $SU(4)$ random vortex world-surface model.


\begin{thebibliography}{99}
\bibitem{hooft} G.'t~Hooft,
\emph{On the phase transition towards permanent quark confinement},
\emph{Nucl.~Phys.} {\bf B138} (1978) 1.
\bibitem{mack} G.~Mack, \emph{Predictions of a theory of quark confinement},
\emph{Phys.~Rev.~Lett.} {\bf 45} (1980) 1378.
\bibitem{olesen} H.~B.~Nielsen and P.~Olesen,
\emph{A quantum liquid model for the QCD vacuum: Gauge and rotational
invariance of domained and quantized homogeneous color fields},
\emph{Nucl.~Phys.} {\bf B160} (1979) 380.
\bibitem{jeff} L.~Del~Debbio, M.~Faber, J.~Giedt, J.~Greensite and
\v{S}.~Olejn{\'\i}k, \emph{Detection of center vortices in the lattice
Yang-Mills vacuum}, \emph{Phys.~Rev.~D} {\bf 58} (1998) 094501
[{\tt hep-lat/9801027}].
\bibitem{percol} M.~Engelhardt, K.~Langfeld, H.~Reinhardt and O.~Tennert,
\emph{Deconfinement in $SU(2)$ Yang-Mills theory as a center vortex
percolation transition}, \emph{Phys.~Rev.~D} {\bf 61} (2000) 054504
[{\tt hep-lat/9904004}].
\bibitem{philippe} P.~de~Forcrand and M.~D'Elia,
\emph{On the relevance of center vortices to QCD},
\emph{Phys.~Rev.~Lett.} {\bf 82} (1999) 4582 [{\tt hep-lat/9901020}].
\bibitem{su2conf} M.~Engelhardt and H.~Reinhardt,
\emph{Center vortex model for the infrared sector of Yang-Mills
theory: Confinement and deconfinement}, \emph{Nucl.~Phys.} {\bf B585}
(2000) 591 [{\tt hep-lat/9912003}].
\bibitem{su2top} M.~Engelhardt,
\emph{Center vortex model for the infrared sector of Yang-Mills
theory: Topological susceptibility}, \emph{Nucl.~Phys.} {\bf B585}
(2000) 614 [{\tt hep-lat/0004013}].
\bibitem{su2csb} M.~Engelhardt,
\emph{Center vortex model for the infrared sector of Yang-Mills
theory: Quenched Dirac spectrum and chiral condensate},
\emph{Nucl.~Phys.} {\bf B638} (2002) 81 [{\tt hep-lat/0204002}].
\bibitem{su3conf} M.~Engelhardt, M.~Quandt and H.~Reinhardt,
\emph{Center vortex model for the infrared sector of $SU(3)$ Yang-Mills
theory: Confinement and deconfinement}, \emph{Nucl.~Phys.} {\bf B685}
(2004) 227 [{\tt hep-lat/0311029}].
\bibitem{su3bary} M.~Engelhardt,
\emph{Center vortex model for the infrared sector of $SU(3)$ Yang-Mills
theory: Baryonic potential}, \emph{Phys.~Rev.~D} {\bf 70} (2004) 074004
[{\tt hep-lat/0406022}].
\bibitem{su3freee} M.~Quandt, H.~Reinhardt and M.~Engelhardt,
\emph{Center vortex model for the infrared sector of $SU(3)$ Yang-Mills
theory: Vortex free energy}, \emph{Phys.~Rev.~D} {\bf 71} (2005) 054026
[{\tt hep-lat/0412033}].
\bibitem{jeffstef} J.~Greensite and \v{S}.~Olejn{\'\i}k,
\emph{$k$-string tensions and center vortices at large $N$},
\emph{JHEP} {\bf 0209} (2002) 039 [{\tt hep-lat/0209088}].
\bibitem{cornwall} J.~M.~Cornwall,
\emph{Nexus solitons in the center vortex picture of QCD},
\emph{Phys.~Rev.~D} {\bf 58} (1998) 105028 [{\tt hep-th/9806007}].
\bibitem{contvort} M.~Engelhardt and H.~Reinhardt,
\emph{Center projection vortices in continuum Yang-Mills theory},
\emph{Nucl.~Phys.} {\bf B567} (2000) 249 [{\tt hep-th/9907139}].
\bibitem{luctep03} B.~Lucini, M.~Teper and U.~Wenger,
\emph{The high temperature phase transition in $SU(N)$ gauge theories},
\emph{JHEP} {\bf 0401} (2004) 061 [{\tt hep-lat/0307017}].
\bibitem{luctep04} B.~Lucini, M.~Teper and U.~Wenger,
\emph{Glueballs and $k$-strings in $SU(N)$ gauge theories: Calculations with
improved operators}, \emph{JHEP} {\bf 0406} (2004) 012
[{\tt hep-lat/0404008}].
\bibitem{luctepnew} B.~Lucini, M.~Teper and U.~Wenger,
\emph{Properties of the deconfining phase transition in $SU(N)$ gauge
theories}, \emph{JHEP} {\bf 0502} (2005) 033 [{\tt hep-lat/0502003}].
\end{thebibliography}
\end{document}